\documentclass[10pt, a4paper]{article}
 \usepackage{graphicx}                    
 \usepackage{color}                       
 \usepackage{url}                         
 
\begin{document}



Title: To the question about perturbations of solar-terrestrial characteristics.

Authors: E. A. Gavryuseva (Institute for Nuclear Research RAS)

Comments: 14 pages, 6 Postscript figures

\begin{abstract}
%
Data obtained over the last three solar cycles have been analysed
to reveal the relationships
between theintensity of the photospheric field measured along the line of sight
by the WSO group at heliolatitudes from -75 to 75 degrees 
and the intensity of the interplanetary magnetic field and absolute values 
of the perturbations of the different characteristics
of the solar wind at the Earth orbit, and geomagnetic parameters.
provided by the OMNI team.

 The heliospheric and geomagnetic data are found to be
 divided into two groups characterized by their response to variability of the
 solar magnetic field latitudinal structures on short and on long time scales.
\end{abstract}
\vspace{1pc}
\section{Introduction}

Solar magnetic field plays the main role in the heliosphere. 
It has to be study carefully the relations between the solar wind, 
the perturbations of geosphere and global structure of the photosperic magnetic field.
WSO and OMNI data were used for comparison of 30 years long data sets 
(see referenses in Gavryuseva 2018c) on a long and short time scale. 
This paper is concentrated on the study the relations of between absolute values of these characteristics. 

    This approach led to the conclusion that all solar wind data
  and geomagnetic perturbations that were examined divided into two groups
  characterized by sensitivity to the variability of the interplanetary
  magnetic field  and photospheric field at different latitudes.

  We use  Wilcox Solar Observatory  data
 for the photospheric magnetic fields
 http://wso.stanford.edu/synopticl.html,
  (Scherrer et al.,  1977),
 OMNI data for solar wind parameters at the
 Earth's orbit, and indices of geomagnetic activity
 for the period 1976-2004,
 to study the relations between the solar wind,
 geomagnetic disturbances and
 solar drivers at different solar latitudes.

  In order to understand from which latitudinal zone
  the solar wind is originated and
  how it depends on the activity cycle
  it is necessary to know the latitudinal $SMF$ 
  structure over at least 22 years.

 The latitudinal structure of the $SMF$ has been deduced
 for the last 29 years since May 27, 1976
 from the Wilcox Solar Observatory (WSO) data
  (Scherrer et al.,  1977;
   Gavryuseva \& Kroussanova,  2003, Gavryuseva \& Gogoli, 2006, Gavryuseva, 2005, 2006, 2006a,b, 2008a,b, 2018 and references there).
 The structure in latitude and time of the 1-year running mean
 of the solar magnetic field
 with 1 Bartels Rotation (BR, 1 BR = 27 days) step is shown 
 on the upper plot in Fig. 1 Gavryuseva, 2018c.

    The solar wind and geomagnetic  data were taken from the OMNI directory
  (http://nssdc.gsfc.nasa.gov/omniweb)
  which contains the Bartels mean values of the interplanetary magnetic  field 
 (IMF)
  and solar wind plasma parameters measured by various
  space-crafts near  the  Earth's  orbit,  as  well  as  geomagnetic
  and solar   activity indices).
  First, daily averages are deduced from OMNI's basic hourly values,
  and then the 27-day Bartels averages are deduced from 
  the daily averages.
  The corresponding standard deviations are related to only these averages
  and do not include the variances in the higher resolution data.


The $IMF$ and solar wind parameters taken into account
 are the following:\\
  $B_x$,   $B_y$,  $B_z$ and
  $B = (B_x^2+B_y^2+B_z^2)^{1/2}$
  are the components and magnitude
  of the interplanetary magnetic field, in nT;\\
  Proton density, $N_p$,   in $N/cm^3$;\\
  Proton temperature, $T_p$, in degrees  $K$;\\
  Plasma speed, $V_p$, in $km/s$;\\
  Electric field, in mV/m;\\
  Plasma beta,
   $N_{\beta}= [(T*4.16/10^5) + 5.34] * N_p/B^2$;\\
  Ratio   $N_{\alpha}/N_{p}$;\\
   Flow Pressure, $P$ proportional to $N_p*V^2$, in nPa;\\
   Alfven Mach number,      $M_a = (V*N_p^{0.5})/20*B$.\\
  The geomagnetic parameters taken into account are the following:\\
  $AE$-index;\\
  Planetary Geomagnetic Activity Index, $K_p-$ index;\\
  $DST$-index, in nT.\\
  Sunspot number ($SSN$) was used, as well, for a further comparison.

The $X$ axis directed along the intersection line of the ecliptic and solar equatorial 
planes to the Sun, the Z axis is directed perpendicular 
and north-ward from the solar equator, 
and the $Y$ axis completes the right-handed set. 

  The solar wind parameters
  analysed cover the same period as the WSO solar data
  with one  Bartels rotation  resolution.
  We call the set of these 16 parameters taken
  from the OMNI data base as "solar wind"
  ($SW$) data; they include the
  interplanetary magnetic field,
  solar wind and geomagnetic parameters and 
  sun spot number ($SSN$).
   
  \section{Relationships between the Solar Magnetic Field Intensity/ and 
   the Absolute Values of the OMNI Data/  on Long- and Short-Term Scales}
   Physical connections between the Sun and
   the  interplanetary parameters
   could be attributed to the influence of the
   intensity of the solar magnetic field
   without taking into account its polarity ($|MF|$).
   The corresponding  correlation coefficients
   between the 1-year mean values of the
   $SMF$ intensity and absolute values of the $SW$ data ($|SW|$)
   as functions of time delay in years
   and in latitude  are  shown in Fig. 1.
   The  $K_{cor}(|MF|, |SW|)$ have an 11-year periodicity
   for all the $SW$ data except the $IMF$ intensity $B$,
   and this periodicity is slightly visible in the $K_{cor}$ for the
   absolute values of the $B_y$: $|B_y|$ and $AE$: $|AE|$.

   There is a remarkable particularity of the latitudinal dependence
   of the\\   $K_{cor}(|MF|, |SW|)$: the change of 
   sign and
   a phase shift of the correlation coefficients at the  
   heliographic latitude
   of about 50-55 degrees. This result can be interpreted as an evidence 
   that the photospheric magnetic fields originated 
   at the heliographic latitudes up to  $\pm 55$ degrees propagate in 
   the heliosphere and contribute to the perturbations of
   the solar wind and magnetosphere. The solar magnetic fields
   originated above $\pm 55$ degrees do not appear close to the Earth's orbit
   (see also  Gavryuseva, 2006c,f; 2008b;  Gavryuseva \& Gondoli, 2006).

   The  $K_{cor}(|MF|, |SW|)$ are symmetric respect to the equator.
   This is well illustrated in Fig. 2 for
   the latitudinal dependence  of the  $K_{cor}(|MF|, |SW|)$
   for the fixed optimum delay between the $SMF$ intensity and
   the absolute values of the $SW$ data.
   Thise figures  are  analogous to Fig. 11, 12  for the original 
   (not absolute) $SMF$ (Gavryuseva, 2018c)
   and $SW$ values where the $K_{cor}(MF, SW)$ are
   antisymmetric to the equator.
   There is a clear anti-correlation between the intensity of the
   photospheric field of the activity belts and $B$, $V_p$, $P$, $N_p$,
   $N_{\beta}$ and $M_a$.
   This confirms that during high activity periods
   the slow solar wind prevails.
   Positive correlations of the $|MF|$
   with the absolute values of the $AE$, $K_p$ and $DST$ indices
   correspond to the statement that most of the geomagnetic 
   perturbations are originated from the low and middle latitudes
   when the intensity of the magnetic field is high.
   Figure 3 
     shows the latitudinal dependence  of the  $K_{cor}$
   for the fixed optimum delay between the absolute values of
   the $SMF$ and $SW$ data after the filtering of the variabilities
   longer than 4 years and shorter than 1 year.
   The $K_{cor}$  for -$B_y$, $T_p$, $V_p$, $P$, $N_p$,
   $AE$, $K_p$ and $-DST$  have very similar latitudinal dependence,
   and the strongest correlation with the $SMF$ intensity takes place
   at about 40 degrees in the southern hemisphere.

   Owing to the symmetry of the solar magnetic field intensity $|MF|$
   the latitudinal dependences
   of the $K_{cor}(|MF|, |SW|)$ and $K_{cor}(F|MF|, F|SW|)$
   are also symmetric  respect to the equator.
   These correlation coefficients are plotted in Fig. 4
  together for all the $|SW|$ data
   for the long-term variability (on the left side)
   and for the short-term variability (on the right side).
   Figure 4 for the $K_{cor}$
    of the absolute values of the $SMF$ and $SW$ data
    is analogous to the corresponding  Figs. 14 and 15
    for the original values of the $SMF$ and $SW$ data (Gavryuseva, 2018c).
    The difference between them permits to investigate
    the sensitivity of the $SW$ parameters to the $SMF$ polarity
    (and not only to the intensity of the SMF) and
    to the basic topology of the solar magnetic field.
 \section{Cross-Correlation between the OMNI Data}
        The cross-correlation coefficients between the OMNI data  provide 
    an information limited to the relationships between them and could be 
    useful for understanding why the $SW$ parameters are subdivided into  two groups only.

    The parameters  of the interplanetary field,
     solar wind and geomagnetic activity  have been numbered as follows:\\
    1:~$B$,   \ \  2:~$B_x$, \ \    3:~$B_y$, \ \   4:~$B_z$,\ \
    5:~$T_p$, \ \  6:~$V_p$, \ \    7:~$E$,   \ \   8:~$N_{\alpha}/N_p$,\ \
    9:~$N_p$, \ \  10:~$P$,  \ \   11:~$N_{\beta}$, \ \  12:~$M_a$,\ \
   13:~$AE$,  \ \  14:~$K_p$, \ \  15:~$DST$, \ \  16:~$SSN$.\\
    The cross-correlation was calculated and the corresponding
    correlation coefficients are plotted in  Fig. 5.
    The corresponding numbers and short names of the parameters
    are shown along the $X$  and $Y$ axis.
    Asterisks correspond to the positive values of $K_{cor}$,
    and diamonds correspond to the negative values of $K_{cor}$.
    The size of the symbols is proportional to their values.
    On the top there are $K_{cor}$
    for the original values of the $SW$ data (on the left side),
    and for their absolute values (on the right side).
    This makes difference for the $IMF$ components $B_x$, $B_y$ and $B_z$,
    for $E$ and for $DST$.
    On the bottom  there are $K_{cor}$ for the residuals of the original values 
    of the $FSW$ data (on the left side),
    and for the residuals of their absolute values (on the right side).

     From the  first plot on the left top   strong correlations
    between the parameters No 9, 10, 11, 12, 13 
    and between  No 5, 6, 14    are  deduced.
    Not very strong, but significant correlations
    take place between   No 5, 6, 9, 13. 
    Strong anti-correlations take place between
    No 4 and 7; No 14 and 15; No 5, 6 and 15, etc.
    Then there are anti-correlations between
    No 8 and 9, 10, 11, 12, 13, etc.

    Comparison between the other plots of Fig. 5   
    and the groups
    of the $SW$ data connected with the $B$ and $B_z$ (or $B_x$, $B_y$)
    permits to verify and to understand the existence of such groups.
    The study of such cross-relationships is very informative,
    but it does not provide
    a sufficient support to predict the $SW$ connection with the
    solar magnetic field while  it
    helps to understand which of $SW$ parameters  could respond
    in a similar way to the solar perturbations.

    Direct comparison of the $SW$ data
    with the basic topology of the $SMF$ 
    permits to study SMF -- SW relationships.

  An influence of the solar activity 
  (or solar magnetic field intensity)
  on the $SW$ parameters was investigated by the
  analysis of their correlation on short subsequent
  intervals of time.
  Fig. 6 
  shows the correlation through  solar cycles  between
  the 3-year long sub-sets of the data
  corresponding to the absolute values
  of the photospheric field and to the
  absolute values  of
  the  solar wind and geomagnetic parameters.
  It is clearly visible that
  the main source of geomagnetic perturbations is concentrated
  in the helio-latitudinal zone from -55 to 55 degrees about  
   \section{Some summary remarks}
  Southward-directed interplanetary magnetic field is considered
  a primary cause of geomagnetic perturbations
  (Durney, 1961;  Gonzales et al., 1994, 1999).
  As  a consequence the orientation of the interplanetary magnetic field
  (Axford and McKenzie, 1997;  Low, 1996;
    Parker, 1997;  Smith, 1997)
  plays an important role.

  The solar activity phenomena depend on the sunspot cycle, which can be
  characterized by the variability of the $SMF$ intensity
  in time and along the latitudes.  The topology of the solar
  magnetic field influences the geomagnetic perturbations
  through the intensity and orientation of the interplanetary magnetic
  field and/or through other parameters of the solar wind.
  In this approach we could understand the presence of two groups of
  the OMNI data similarly sensitive to the basic topology of the magnetic
  field of the Sun (from the point of view of the dependence on latitude 
  and phase-shift of the correlation of the coefficients
  with the mean latitudinal magnetic field).

  The formal and complete study of the problem of solar-terrestrial relations 
has been performed and the connections between the processes on the way
 from the Sun to the Earth have been revealed. A useful information was 
deduced from the temporal behaviour and dependence of the correlation 
of the photospheric magnetic field and different parameters of interplanetary
 space and geomagnetosphere.

It was revealed directly from the experimental data that there are 
two groups of $SW$ parameter which respond in a similar way to
the behaviour of solar characteristics.
  We found that the photospheric field
  influences  the magnitude of the interplanetary field
  and, in the same way, the proton density, flow pressure,
  Alfven Mach number and plasma $\beta$ respond to the $SMF$.
  Moreover the $AE$-index behaves in a similar way as the above mentioned
  solar wind parameters.

  On the contrary, regarding  the planetary geomagnetic activity index $K_p$
  we can deduce that solar activity events  (CME, magnetic field intensity, 
sunspots, etc.)
  through perturbations of
  the $B_z$ component ($B_x$, $B_y$ components) of the $IMF$, 
  the   proton temperature $T_p$, plasma  speed $V_p$,
  $N_{\alpha}/N_{p}$ ratio influence the $K_p$ index.
  The variations of the $-B_z$ ($B_y$) component
  produce the perturbations
  of the $DST$ index, and they are of opposite sign of  the $K_p$
  and $B_x$ time   dependence.

   It was also revealed  from the experimental data that the solar magnetic fields and 
solar activity  processes  originated bellow $\pm 55$ degrees  propagate up  to the 
Earth orbit and produce the perturbations of the magnetosphere
   (Gavryuseva, 2006 c,f; 2008b,  Gavryuseva \& Godoli, 2006).

 These results are useful for understanding
 the origin of solar wind and geomagnetic perturbations
 and for long-term predictions.

 \section*{Acknowledgments}
      I thank the WSO and OMNI teams for making available
      data of measurements of the solar magnetic field,
      solar wind and geomagnetic quantities.
    I am grateful to Prof. G. Godoli  for his stimulating interest in these results
    and Profs. B. Draine, L. Paterno and E. Tikhomolov  for  help in polishing this paper 
    and useful advises.

\newpage
\clearpage
\clearpage
 \begin{figure}
  \centerline{
  \includegraphics[angle=90, width=39pc]
    {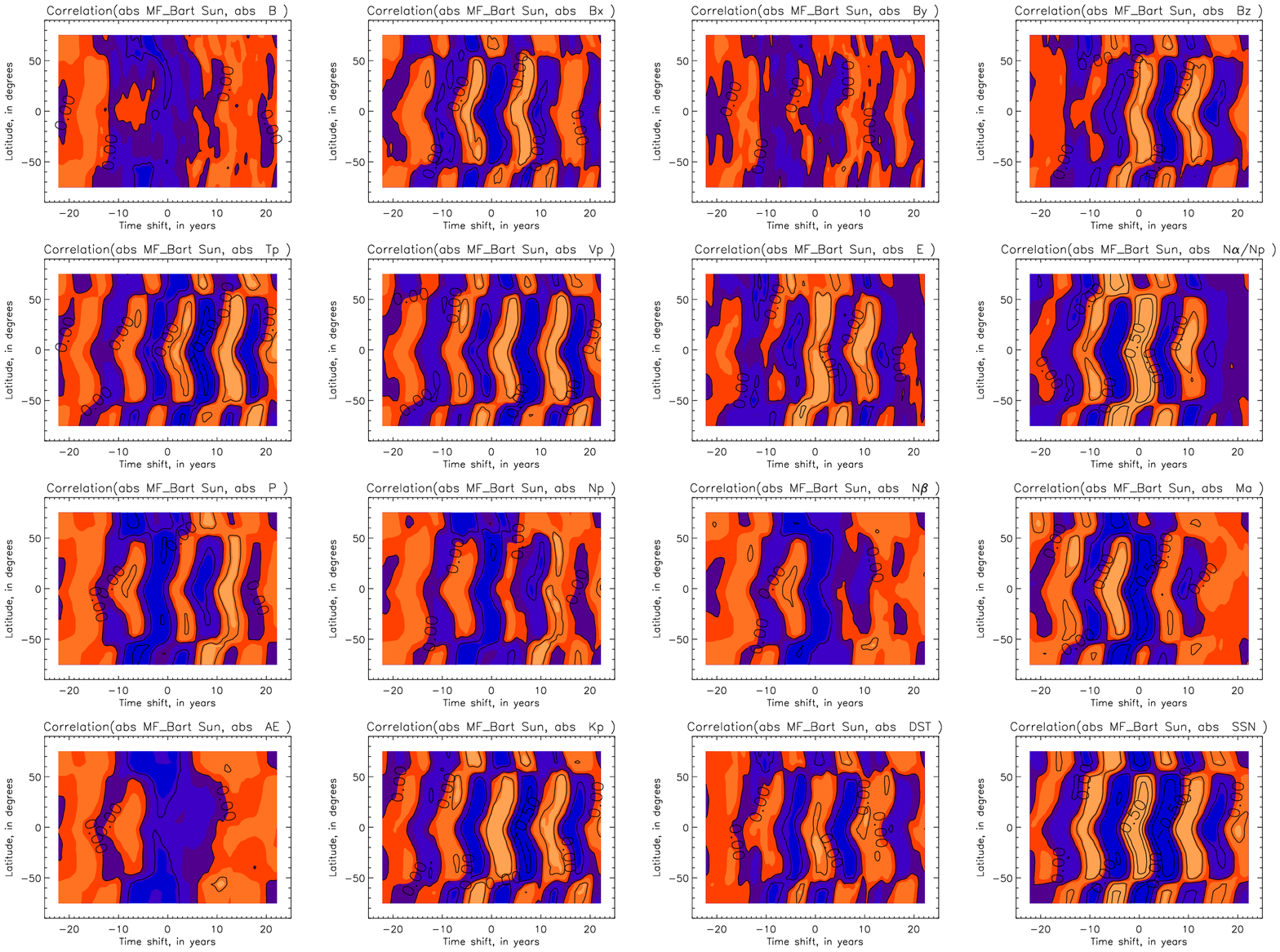}
}
\caption{
 The correlation coefficients of  absolute values of the photospheric field
 at different latitudes ($Y$ axis)
 with absolute values of solar wind and  geomagnetic parameters
 as a function of delay and latitude.
 Orange and red (blue) colors indicate
 positive (negative) correlation coefficient values.
 }
  \end{figure}
\clearpage
 \begin{figure}
  \centerline{
  \includegraphics[angle=90, width=39pc]
    {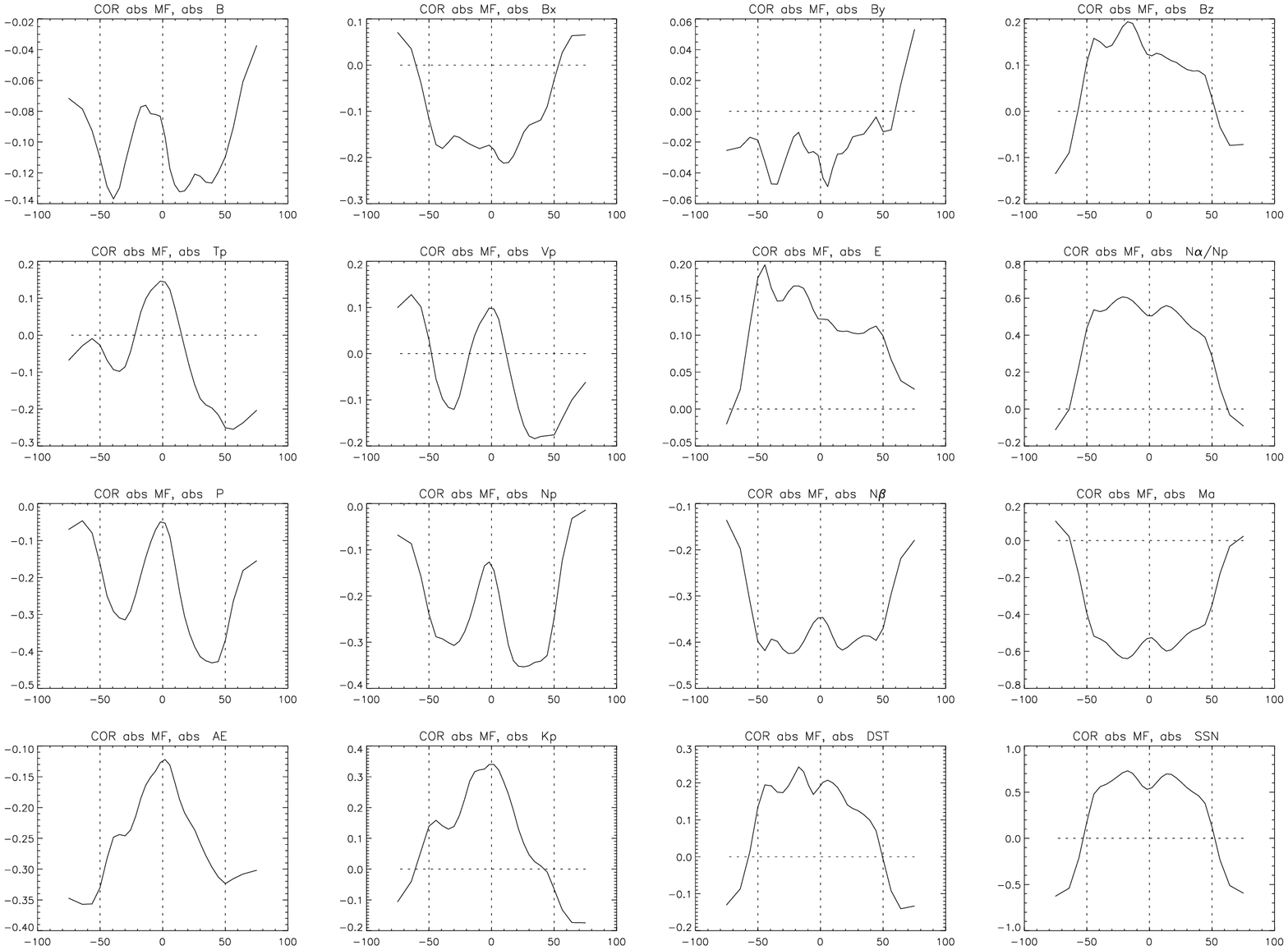}
  }
  \caption{
  Coefficients of correlation between  1-year mean
  of the  absolute value of the photospheric field
  at different latitudes and
  the interplanetary magnetic field,
  solar wind and geomagnetic parameters with the fixed delay 
  of 4 days as a function of latitude ($X$ axis).
 }
  \end{figure}
\clearpage
 \begin{figure}
  \centerline{
  \includegraphics[angle=90, width=39pc]
   {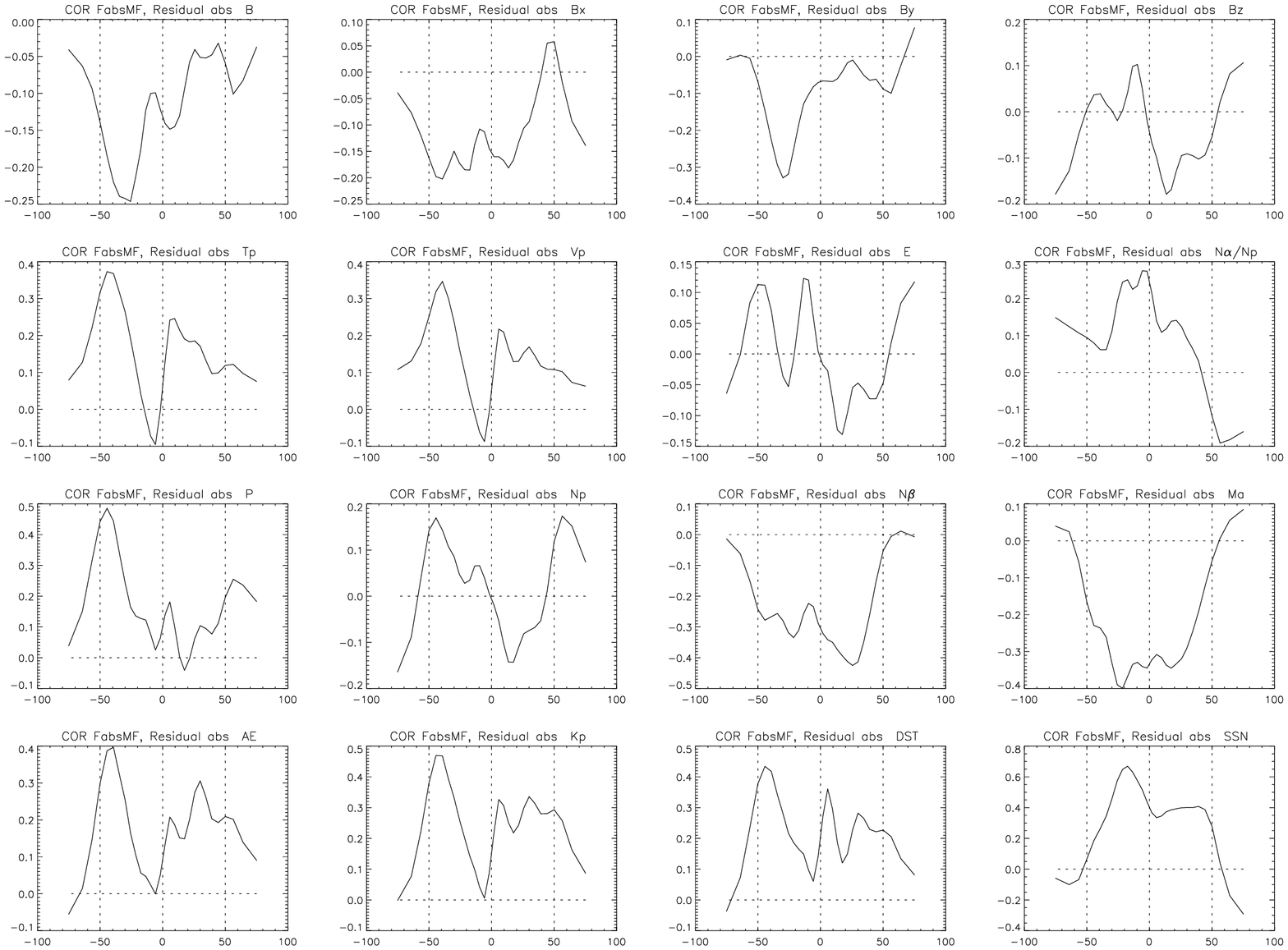}
  }
  \caption{
  Coefficient of correlation between  short term variabilities
  of the absolute value of the photospheric field
  at different latitudes and
  the interplanetary magnetic field,
  solar wind and geomagnetic parameters with the fixed delay 
  of 4 days as a function of latitude ($X$ axis).
 }
  \end{figure}
\clearpage
\begin{figure}
 \centerline{
 \includegraphics[angle=90, width=39pc]
   {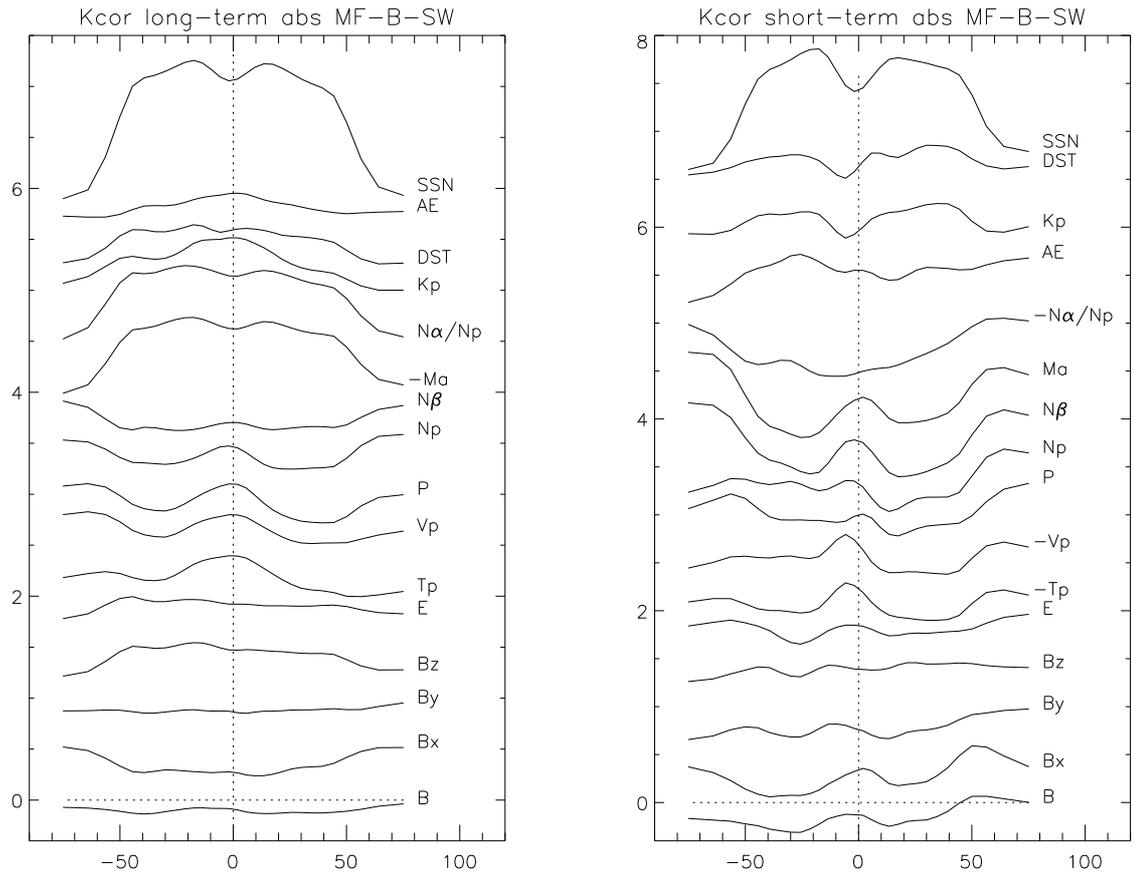}
  }
  \caption{
 Coefficients of correlation between 
 the absolute value of 
 the photospheric field and
 interplanetary magnetic field,
 solar wind and geomagnetic parameters
 (marked on the right end of the corresponding curve)
 for the mean values over 1 year (on the left plot)
 and for the short term variable  part of them (on the right plot)
 at different latitudes with a fixed delay of 4 days.
}
 \end{figure}
\clearpage
 \begin{figure}
  \centerline{
  \includegraphics[angle=90, width=39pc]
    {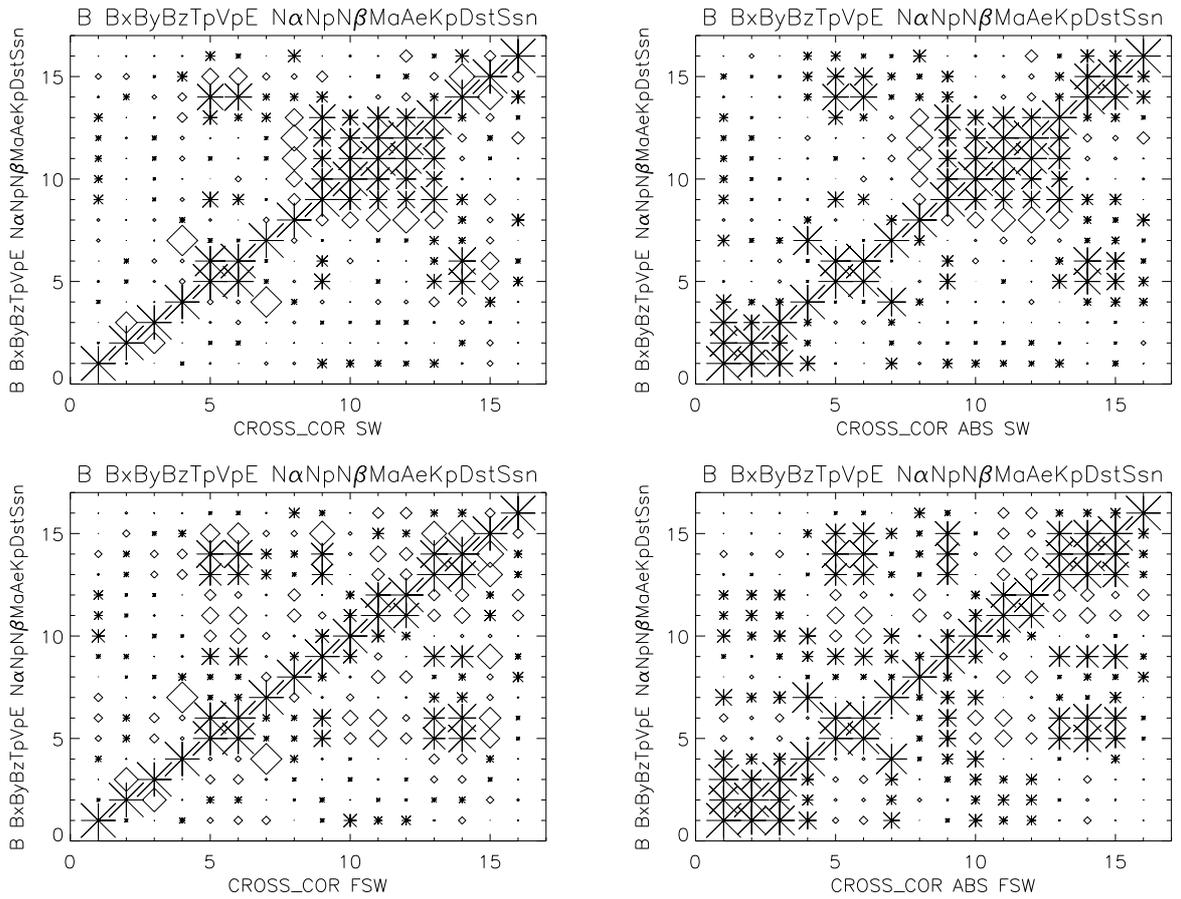}
  }
  \caption{
  Coefficients of correlations between the  yearly means of the
  intensity, original values of the components of
  the interplanetary magnetic field,
  solar wind and geomagnetic parameters
  (marked by number in $X$ and $Y$ axis)
  are shown on the upper plot on the left side.
  The same for their absolute values is shown on the top on the right site.
  Coefficients of correlation for short-term variability of the
  the intensity, components of the $IMF$, solar wind and geomagnetic
  parameters  are shown on the bottom plot on the left.
  The same for the short-term variability of the absolute values of the $SW$ data
   is shown on the bottom plot on  the right.
 }
  \end{figure}
\clearpage
  \begin{figure}
  \centerline{
   \includegraphics[angle=90, width=39pc]
    {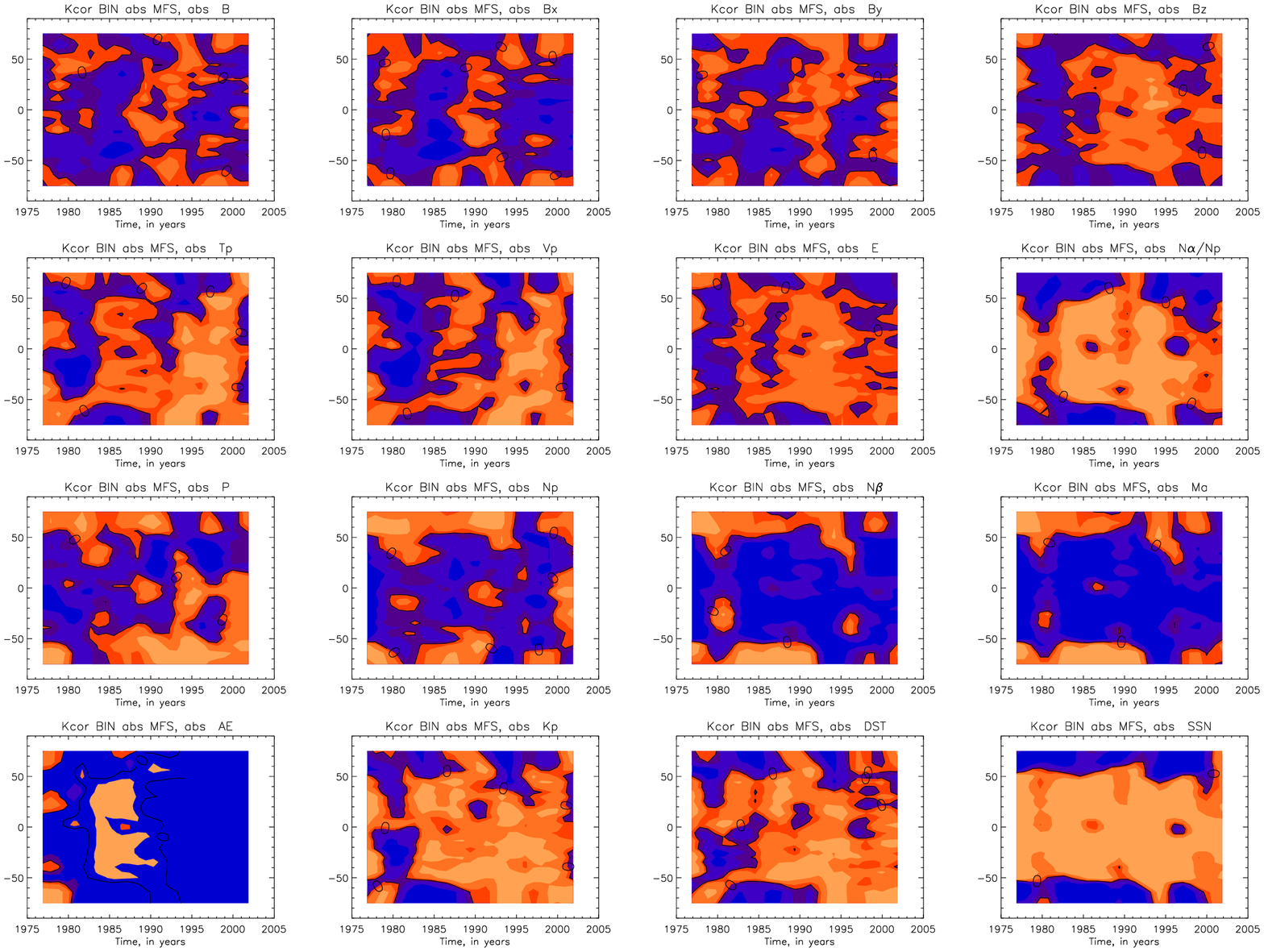}
}    
\caption{
   The correlation coefficient through  solar cycles between
   the 3-year long  sub-sets of data
   of the photospheric field intensity
   at different latitudes ($Y$ axis)
   with the  absolute values of the solar wind
   and geomagnetic parameters.
   Orange and red (blue) colors indicate
   positive (negative) correlation coefficient values.
  }
  \end{figure}

\begin{thebibliography}{}
  \item
  Axford, W.L. and McKenzie J.F. (1997),
  The solar wind,
  in Cosmic winds and the heliosphere,
   ed. by J.R. Jokipii, C.P. Sonett \& M.S. Giampapa,
    31.  

\item
Clua de Gonzalez, W. D. Gonzalez, S. L. G. Dutra, B. T. Tsurutani (1993),
Periodic Variation in the Geomagnetic Activity: A Study Based on the Ap Index,
J. Geophys. Res., 98, 9215.

  \item
  Durney, J.W.   (1961),
  Interplanetary magnetic field and the auroral zones,
   Phys. Rev. Lett., 6, 47.  

\item
Fraser-Smith, A. C. (1973), 
Solar cycle control in the 27-day variation of geomagnetic activity,
J. Geophys. Res., 78, 5825.

\item
   Gavryuseva, E. (2005), Latitudinal streams of solar magnetic field,
    Proc. of 11 Int. Scientific Conf. Solar-Terrestrial Influences,  Nov. 2005,
    BAS, 229-233. 

  \item
  Gavryuseva, E. (2006),
  Topology and dynamics of the  magnetic field
  of the Sun,
   News of the Academy of Science, IzvRAN, ser. Physics,
   70, No. 1, 102.  

\item
  Gavryuseva, E. (2006a),
Latitudinal Structure of the Photospheric Magnetic Field through solar cycles
Solar Activity and its Magnetic Origin, Proc. of the 233rd Symposium of the
IAU, Cairo, Egypt, March 31 - April 4, 2006,
Ed. V. Bothmer; A. A. Hady. Cambridge:
       Cambridge University Press, 124.  

\item
  Gavryuseva, E. (2006b),
Basic topology and dynamics of magnetic field leading activity the Sun
Solar Activity and its Magnetic Origin, Proc. of the 233rd Symposium of the
IAU, Cairo, Egypt, March 31 - April 4, 2006,
Ed. V. Bothmer; A. A. Hady. Cambridge:
       Cambridge University Press, 67.  

\item
  Gavryuseva, E. (2006c),
Variability of the differential rotation of the photospheric magnetic field through solar cycles
Solar Activity and its Magnetic Origin, Proc. of the 233rd Symposium of the
IAU, Cairo, Egypt, March 31 - April 4, 2006,
Ed. V. Bothmer; A. A. Hady. Cambridge:
       Cambridge University Press, 65.  

\item
  Gavryuseva, E. (2006d),
North-South asymmetry of the photospheric magnetic field
Solar Activity and its Magnetic Origin, Proc. of the 233rd Symposium of the
IAU, Cairo, Egypt, March 31 - April 4, 2006,
Ed. V. Bothmer; A. A. Hady. Cambridge:
       Cambridge University Press, 63.  

\item
  Gavryuseva, E. (2006e),
Longitudinal structure of the photospheric magnetic field
Solar Activity and its Magnetic Origin, Proc. of the 233rd Symposium of the
IAU, Cairo, Egypt, March 31 - April 4, 2006,
Ed. V. Bothmer; A. A. Hady. Cambridge:
       Cambridge University Press, 61.  

\item
  Gavryuseva, E. (2006f),
Relationships between photospheric magnetic field, solar wind and geomagnetic perturbations
over last 30 years
Solar Activity and its Magnetic Origin, Proc. of the 233rd Symposium of the
IAU, Cairo, Egypt, March 31 - April 4, 2006,
Ed. V. Bothmer; A. A. Hady. Cambridge:
       Cambridge University Press, 291.  

\item
  Gavryuseva, E. (2008a),
  In search of the origin of the latitudinal structure of the photospheric magnetic field,
  ASP Conf. Ser., v. 383,
  Proc. of "Subsurface and atmospheric influence on solar activity", held at
  NSO, Sacramento Peak, Sunspot, New Mexico, USA
  16-20 April 2007,
  Ed. R. Howe, R. W. Komm, K. S. Balasubramaniam \& G. J. D. Petrie,
  99.  

\item
  Gavryuseva, E. (2008b),
  Longitudinal structure originated in the tachocline zone of the Sun,
  ASP Conf. Ser., v. 383,
  Proc. of "Subsurface and atmospheric influence on solar activity", held at
  NSO, Sacramento Peak, Sunspot, New Mexico, USA
  16-20 April 2007,
  Ed. R. Howe, R. W. Komm, K. S. Balasubramaniam \& G. J. D. Petrie,
  381.  

\item
   Gavryuseva, E. (2018),
     Latitudinal structure and dynamic of the photospheric magnetic field,
    arXiv:1802.02450.   
\item
    Relations between variability of solar and interplanetary characteristics
\item
 To the connection between intensity of the solar and geomagnetic perturbations,
   arXiv:1802.NNNN(N) 
\item
   Gavryuseva, E. (2018),
     Latitudinal structure and dynamic of the photospheric magnetic field,
    arXiv:1802.02450
\item
   Gavryuseva, E. (2018b),
   Longitudinal structure of the photospheric magnetic field in Carrington system,
   arXiv:1802.02461   

\item
Gavryuseva, E.; Godoli, G.  (2006),
Structure and rotation of the large scale solar magnetic field
 observed at the Wilcox Solar Observatory
Physics and Chemistry of the Earth, v. 31, issue 1-3,  68.  

\item
  Gavryuseva, E., Kroussanova, N. (2003),
Topology and dynamics of the Sun's magnetic field
SOLAR WIND TEN: Proceedings of the Tenth International Solar Wind Conference,
AIP Conference Proceedings, v. 679,  242.  

\item
    Gavryuseva, E., and  V. Gavryusev (1994),
      Time variations of the $^{37}Ar$ production rate
      in chlorine solar neutrino experiment,
 Astron. Astrophys, 283, 978.  
  
\item
   Gavryuseva, E., and V. Gavryusev (2000),
   Solar variability and its prediction,
   Long and short term variability
   in Sun's history and global change,
   ed. W.Schroder, Science Edition, Bremen, Germany, p.89.  

 \item
   Gavryuseva, E., and N. Kroussanova,  (2003),
   Topology and dynamic of solar magnetic field,
   Proc. of the Tenth International Solar Wind Conference, 
    AIP Conf. Proc., v. 679, 242.  

\item
     Gavryusev, V., E. Gavryuseva, Ph. Delache, and F. Laclare (1994),
     Periodicities in solar radius measurements,
     Astron. Astrophys, 286, 305.  

\item
   Gavryuseva, E., V. Gavryusev, and M.P. Di Mauro (2000),
   Internal rotation of the Sun as inferred from GONG observations,
   Astronomy Lett., 26, N 4, 261.
 \item
    Gavryuseva, E., \& G. Godoli (2006),
    Structure and rotation of the large scale solar
    magnetic field observed at the Wilcox Solar Observatory,
     Physics and Chemistry of the Earth, Elsevier, 
     31, 68. 
 \item
   Gonzalez, W.D., J.A. Joselyn, Y. Kamide, H.W. Krorhl, G. Rostoker,
   B.T. Tsurutani and V.M. Vasyliunas (1994),
   What is a geomagnetic storm?,
   J. Geophys. Res., 99, 5771.  

 \item
   Gonzalez, W.D., B.T. Tsurutani and A.L. Clua de Gonzalez (1999),
   Interplanetary origin of geomagnetic storms,
   Space Science Rev., 88,
   529.  

 \item
   Kane, R.P. (2005a),
   Difference in the quasi-biennial oscillation 
   and quasi-triennial oscillation characteristics of the solar,
   interplanetary, and terrestrial parameters,
   J. Geophys. Res., 110, A01108.  

 \item
   Kane, R.P. (2005b),
   Short-term periodicities in solar indices,
   Solar Phys., 227, 155.  

 \item
   Li, Y., Luhmann, J. G., Arge, C. N., Ulrich, R., How do solar magnetic fields 
   influence the long term changes of some geomagnetic indexes?, American
   Geophysical  Union, Spring Meeting 2001, abstract SH52A-02, 2001.  

 \item
   Pizzo, V. J., A three-dimensional model of corotating streams in the solar
   wind. III Magnetohydrodynamic streams, J. Geophys. Res., 87, 4374, 1982.  

 \item
   Wang, Y.-M., J. Lean, and N. R. Sheeley, The long-term variation of the 
   Sun's open magnetic flux, Geophys. Res. Lett., 27, 505, 2000.  

 \item
   Low, B.C.  (1996),
   Solar activity and the corona,
   Solar Phys., 167, 217.  

 \item
   Luhmann, J. G., Li, Y., Arge, C. N., Gazis, P. R., Ulrich, R., Solar cycle
   changes in coronal holes and space weather cycles, J. Geophys. Res., 107(A8),
   1154, pp. SMP 3-1, 2002.  

 \item
   Makarov, V. I., Tlatov, A. G., Callebaut, D. K., Obridko, V. N., Increase of 
   the Magnetic Flux From Polar Zones of the sun in the Last 120 Years, Solar
   Physics, v. 206, Issue 2, p. 383-399 (2002).  

 \item
   Parker, E.N., (1997),
   Mass ejection and a brief history of the solar wind concept,
   in Cosmic winds and the heliosphere,
   ed. by J.R. Jokipii, C.P. Sonett and M.S. Giampapa,
   p.3.  

\item
   Rivin, Yu.R. (1989),
   Cycles of the Earth and of the Sun,
   Nauka, IZMIRAN, p.36.  

 \item
   Smith, E.J., (1997),
   Solar wind magnetic field,
   in Cosmic winds and the heliosphere,
   ed. by J.R. Jokipii, C.P. Sonett and M.S. Giampapa, p.425.  

 \item
    Stamper E.J., (1997),
    Solar wind magnetic field,
    in Cosmic winds and the heliosphere,
    ed. by J.R. Jokipii, C.P. Sonett and M.S. Giampapa, p.425.  

 \item
    Stamper, R., Lockwood, Wild, M.N., Clark, T.D.G., (1999),
    Solar causes of the long-term increase of the geomagnetic
    activity, J. Geophys. Res., 104, Issue A12, pp.28,325.

\item
Pizzo, V. J. (1982),
    A three-dimensional model of corotating streams in the solar wind.
    III Magnetohydrodynamic streams,
    J. Geophys. Res., 87, 4374.

\item
    Scherrer, P.H., J.M. Wilcox, L.Svalgaard,
    T.L. Duvall, Ph.H. Dittmer and E.K. Gustafson (1977),
    The mean magnetic field of the Sun: observations at Stanford.
    Solar Phys., 54, 353.  
%

\item
Wang, Y.-M., J. Lean, and N. R. Sheeley (2000),
  The long-term variation of the Sun's open magnetic flux,
   Geophys. Res. Lett., 27, 505.
 \end{thebibliography}
\end{document}